# Role of ambient humidity underestimated in research on correlation between radioactive decay rates and space weather

S. Pommé, K. Pelczar

*European Commission, Joint Research Centre (JRC), Geel, Belgium*
*Corresponding author:* stefaan.pomme@ec.europa.eu

Arising from: Milián-Sánchez, V., *et al*., Scientific Reports 10.1038/s41598-020-64497-0 (2020).

In recent work, Milián-Sánchez et al. [1-3] observed fluctuations in radioactive decay rate measurement series, and after excluding environmental influences (measured indoors) as root causes, they looked for possible correlations with astrophysical variables. They reported positive or negative correlations with geomagnetic activity (GMA) and cosmic-ray activity (CRA). This assertion is at variance with the most accurate measurements of radioactivity, which support the validity of the exponential-decay law [4-14]. If a causal relationship between 'space weather' and radioactive decay rates were true, it would invalidate the notion of invariable decay constants and have unforeseeable practical and theoretical implications.

In spite of the authors' efforts to investigate possible influences of environmental parameters – such as ambient temperature, pressure and humidity – on the detectors' stability, it turns out that the influence of ambient humidity on the instrumentation has been underestimated. In Fig. 1, the measured decay rates of a $^{226}$Ra source with a Geiger-Müller counter [1,3] is shown together with measured ambient humidity data in a weather station in the Valencia region [15]. There appears to be a positive correlation between the humidity and the observed decay rates. This is even more striking when applying a toy physical model in which the humidity data are accumulated as moisture in the set-up [16], shown as a line overlaid on the decay rates in Fig. 1.

The moisture model has also been applied to the other decay rate measurements, the background measurements, as well as the additional tests performed on the capacitance of the detector cable and a reference capacitance [16]. On a qualitative level, positive correlations were observed in every case. Six graphs covering a measurement period of at least 1.9 days and showing distinct variations with time have on average a correlation factor of 0.60 (ranging between 0.35 and 0.8) with the humidity model. It seems fair to conclude that ambient humidity, or correlated weather conditions (such as temperature), must have played a significant role in the measurements in spite of measures taken to exclude such interferences. Consequently, the measurement data do not reflect physical changes in the decay constants caused by 'space weather', but a difference in response of the electronics to changing 'terrestrial weather' conditions.

In conclusion, the work in Refs. [1-3] may be regarded as exploratory research, and the sought-after correlation with space weather as highly speculative. With the discovery of hitherto unnoticed correlations with terrestrial weather conditions, the hypothesised causal correlation between space weather and radioactive decay turns out to be unsubstantiated by evidence and remains unlikely to be true.

## Author contributions

S. Pommé: Concept of work, Interpretation of data, Draft of manuscript

K. Pelczar: Creation of new software, Interpretation of data, Revision of manuscript

## Competing interests

The author(s) declare no competing interests.

Figure Caption

Fig. 1: Relative humidity in the Valencia region (top) in the period 5-9 Dec 2014 [11]. Qualitative comparison (bottom) of measured $^{226}$Ra decay rates [1-3] with a simple moisture accumulation model [12].

Fig. 1

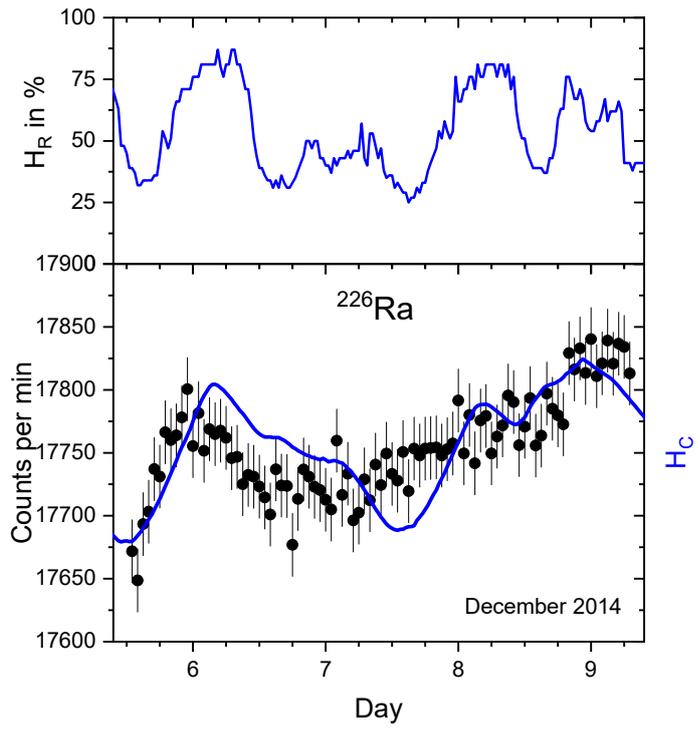